\documentclass[aps,showpacs,prl,twocolumn,footinbib]{revtex4-1}
\usepackage{epsfig}
\usepackage{amsfonts}
\usepackage{amsmath}
\usepackage{bm}
\usepackage{xcolor}
\usepackage{newtxtext,newtxmath}

\usepackage[utf8]{inputenc}
\usepackage[colorlinks,bookmarks=false,citecolor=blue,linkcolor=blue,urlcolor=blue]{hyperref}
\usepackage{soul}
\usepackage{stackengine} 

\newcommand{\ket}[1]{\left| #1 \right>}

\newcommand{\av}[1]{\left< #1 \right>}

\newcommand{\sxk}{{\sigma^x_k}}

\newcommand{\szk}{{\sigma^z_k}}
\newcommand{\smmk}{{\sigma^-_k}}
\newcommand{\sppk}{{\sigma^+_k}}

\newcommand{\upstate}{\ket{\uparrow}}
\newcommand{\downstate}{\ket{\downarrow}}

\newcommand{\be}{\begin{equation}}
\newcommand{\ee}{\end{equation}}

\begin{document}

\title{Exploring non-equilibrium phases of the generalized Dicke model with a trapped Rydberg ion quantum simulator}
\author{F. M. Gambetta, I. Lesanovsky, and W. Li}
\affiliation{School of Physics and Astronomy, University of Nottingham, Nottingham, NG7 2RD, United Kingdom and Centre for the Mathematics and Theoretical Physics of Quantum Non-equilibrium Systems, University of Nottingham, Nottingham NG7 2RD, UK}

\date{\today}

\begin{abstract}
Trapped ions are a versatile platform for the investigation of quantum many-body phenomena, in particular for the study of scenarios where long-range interactions are mediated by phonons. Recent experiments have shown that the trapped ion platform can be augmented by exciting high-lying Rydberg states. This introduces controllable state-dependent interactions that are independent from the phonon structure. However, the many-body physics in this newly accessible regime is largely unexplored. We show that this system grants access to generalized Dicke model physics, where dipolar interactions between ions in Rydberg states drastically alter the collective non-equilibrium behavior. We analyze and classify the emerging dynamical phases and identify a host of non-equilibrium signatures such as multi-phase coexistence regions and phonon-lasing regimes. We moreover show how they can be detected and characterized through the fluorescence signal of scattered photons. Our study thus highlights new capabilities of trapped Rydberg ion systems for creating and detecting quantum non-equilibrium phases.
\end{abstract}

\maketitle

\textit{Introduction---} 
In the last decade trapped ions have been established as a promising experimental platform for investigating the behavior of quantum many-body systems, both in and out of equilibrium~\cite{Schneider:2012,Schindler:2013,Johanning:2009}. Long coherence times and controllable phonon-mediated interactions make this system highly versatile and flexible~\cite{Porras:2004,Deng:2005,Lee:2005}. As a consequence, trapped ions have found important application in the fields of quantum information~\cite{Cirac:1995, Schneider:2012,Schindler:2013,Johanning:2009,Haffner:2008}, metrology~\cite{Blatt:2008,Herrmann:2009,Wineland:2011} and quantum thermodynamics~\cite{Rossnagel:2016,vonLindenfels:2018,Maslennikov:2019}. Moreover, they have been successfully employed to simulate a rich variety of spin many-body quantum models~\cite{Porras:2004,Deng:2005,Friedenauer:2008,Kim:2009,Kim:2010,Kim:2011,Islam:2011,Britton:2012} and provided access to new non-equilibrium collective phenomena, such as non-equilibrium phase transitions~\cite{ Zhang:2017,Zhang:2017DPT}, phonon lasing ~\cite{Wallentowitz:1996,Vahala:2009,Kaplan:2009,Knuz:2010,Xie:2013,Genway:2014,Ip:2018} and quantum synchronization~\cite{Hush:2015}. 

Trapped ion quantum simulators can be further enhanced by exciting ions to highly-lying Rydberg states. Such Rydberg ions, which were initially proposed by M\"uller \textit{et al.}~\cite{Muller:2008,Schmidt-Kaler:2011} and recently experimentally realized~\cite{Feldker:2015,Bachor:2016,Higgins:2017,Higgins:2017PRL},
bear the promise to overcome current scalability limitation of trapped ions setups in quantum information applications~\cite{Cirac:1995,Wineland:1997,Hughes:1996}. Furthermore, the exaggerated properties of Rydberg states~\cite{Saffman:2010,Gallagher:1988,Gallagher:2005,Low:2012} permit the realization of fast quantum gates and, more generally, the implementation and simulation of many-body spin models~\cite{Hague:2012,Li:2014,Vitanov:2019}. 
Particularly intricate scenarios emerge when interactions mediated
by phonons compete with state-dependent dipolar forces among Rydberg states~\cite{Muller:2008,Schmidt-Kaler:2011}. This, together with the strong coupling of Rydberg states to vibrational modes~\cite{Li:2012}, sets the stage for a complex non-equilibrium behavior.

\begin{figure}
	\centering
	\includegraphics[width=0.9\columnwidth]{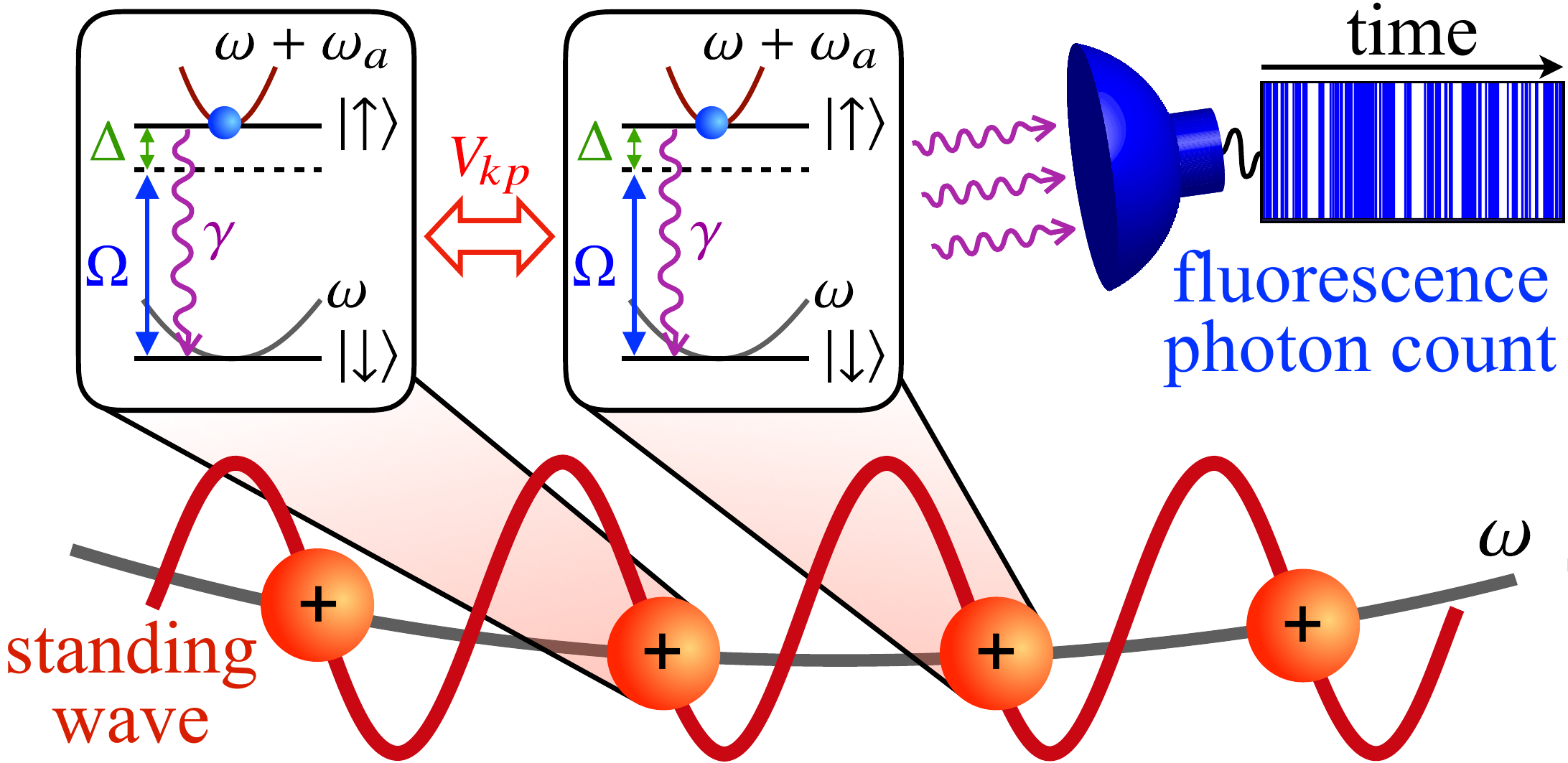}
	\caption{\textbf{Chain of trapped Rydberg ions.} Each ion is modeled as an effective two-level system whose ground state, $ \ket{\downarrow} $, is coupled to a Rydberg excited state, $ \ket{\uparrow} $, by a laser with Rabi frequency $ \Omega $ and detuning $ \Delta $. The state $ \ket{\uparrow} $ spontaneously decays to $ \ket{\downarrow} $ with rate $ \gamma $. Ions at sites $ k $ and $ p $ interact through the interaction potential $ V_{kp} $  and are subject to a state-dependent trapping potential (with trapping frequency $ \omega $ for $ \ket{\downarrow} $ and $ \omega+\omega_a $ for $ \ket{\uparrow} $). The internal states of the ions are also coupled to the phononic degrees of freedom of the chain via a far-detuned standing-wave laser. Aspects of the dynamical behavior of the chain can be probed through the fluorescence signal of emitted photons as a function of time.}
	\label{fig:setup}
\end{figure}
In this work we explore novel non-equilibrium phases that become accessible in trapped ion quantum simulators when this platform is augmented by Rydberg states. We demonstrate that a linear ion chain (see Fig. 1), in which dissipative processes compete with strong coherent interactions, implements an instance of the generalized Dicke model (GDM)~\cite{Genway:2014}. The peculiar properties of Rydberg ions, such as a strong state-dependent coupling between electronic and vibrational degrees of freedom as well as dipolar interactions, give rise to a host of dynamical regimes, including superradiant phases, multiphase coexistence and phonon-lasing (PL) behavior. We show how fingerprints of the different non-equilibrium regimes can be detected in single quantum trajectories~\cite{Dalibard:1992,*Molmer:1993,*Plenio:1998} which, in turn, allow for their experimental observation through time-resolved fluorescence spectroscopy of emitted photons~\cite{Ates:2012,Genway:2014}.

\textit{The model.---} 
The minimal model to describe a trapped Rydberg ion quantum simulator consists of $ N $ two-level effective spin systems, with $  \downstate$ and $ \upstate $ representing the ground and the excited Rydberg state, respectively (see Fig.~\ref{fig:setup}). The two states are coupled by a laser field with Rabi frequency $ \Omega $ and detuning $ \Delta $. The electronic internal states, in turn, are coupled to the vibrational modes of the ion chain through a far-detuned standing-wave laser, which leads to a state-dependent spin-phonon coupling~\cite{GarciaRipoll:2005,Lee:2005,Genway:2014}. For the sake of simplicity, in the following we will consider the presence of the axial center-of-mass (CM) phonon mode only, whose frequency is denoted by $ \omega $. 
The time-evolution of the system density matrix is governed by the quantum master equation (QME)~\cite{BreuerPetruccione}
\begin{equation}\label{eq:QME}
\partial_t \rho(t)=-i[H,\rho]+\mathcal{D}[\rho],
\end{equation}
with Hamiltonian ($ \hbar=1 $)
\begin{align}\label{eq:generalH}
H&=\Omega\sum_{k=1}^{N} \sxk+\Delta \sum_{k=1}^{N} \szk + \sum_{k,p\neq k}V_{kp}\sigma^z_k\sigma^z_p\nonumber\\
&+g\sum_{k=1}^{N} \szk(a^\dagger+a)+\left(\omega +\omega_a \sum_{k=1}^{N} n_k\right)a^\dagger a.
\end{align}
Here, $ \bm{\sigma}_k=(\sigma_k^x,\sigma_k^y,\sigma_k^z) $ are the Pauli matrices acting on the $ k$-th ion, $ n_k=(\mathbb{I}_k+\sigma^z_k)/2 $ is the Rydberg state occupation number operator, and $ a $ ($ a^\dagger $) is the bosonic annihilation (creation) operator of the CM mode. The coupling between internal and vibrational degrees of freedom of the chain is parametrized by $ g $, while $ \omega_a $ measures the difference of the trapping potential between the Rydberg and ground states, respectively~\cite{Muller:2008,Schmidt-Kaler:2011} (see Fig.~\ref{fig:setup}). Here, $ V_{kp}=C_3|\bm{r}_k-\bm{r}_p|^{-3} $ describes a dipole-dipole interaction between an ion at position $ \bm{r}_k $ and one at $ \bm{r}_p $~\cite{Gallagher:2005,Muller:2008,Schmidt-Kaler:2011}. Finally, radiative decay $ \upstate \rightarrow \downstate $ is described through the dissipator
\begin{equation}\label{eq:dissipator} 
\mathcal{D}[\rho]=\gamma\sum_{k=1}^{N}\left[\smmk\rho\sppk-\frac{1}{2}\left\{\sppk \smmk,\rho \right\}\right],
\end{equation}
with $ \sigma^\pm_k=(\sigma^x_k\pm i \sigma^y_k)/2 $.  

To characterize the dynamical behavior of this system we focus, at first, on the mean-field (MF) dynamics of the average displacement, $ X=(A+A^*)/2 $, and momentum, $ P=(A-A^*)/2i $, of the CM mode (with $ A=\av{a} $), and of the average magnetization of the ions, $ \bm{J}=N^{-1}\sum_k\av{\bm{\sigma}_k} $. In terms of these semi-classical variables, the MF equations of motion (EoM) associated with the QME~\eqref{eq:QME} are
\begin{subequations}
	\label{eq:MFEoM}
	\begin{align}
	\dot{X}&=\omega P + \frac{1}{2}\omega_a N(1+J_z) P,\\
	\dot{P}&=g N J_z -\omega X -\frac{1}{2}\omega_a N(1+J_z)X,\\
	\dot{J_x}&=-\mathcal{F}(\bm{J},X,P) J_y-2\mathcal{V}J_zJ_y-\frac{\gamma}{2}J_x,\\
	\dot{J_y}&=\mathcal{F}(\bm{J},X,P) J_x-2\Omega J_z+2\mathcal{V}J_zJ_x-\frac{\gamma}{2}J_y,\\
	\dot{J_z}&=2\Omega J_y-\gamma(1+J_z). 
	\end{align}
\end{subequations}
Here, we have introduced the function $ \mathcal{F}(\bm{J},X,P)=2\big[\Delta+2gX+
\omega_a (X^2+P^2)/2\big] $ and the MF interaction potential $ \mathcal{V}=2 N^{-1}\sum_{k\neq p} V_{kp}$. In order to derive Eq.~\eqref{eq:MFEoM}, we have made the following replacements in the evaluation of expectation values of products of observables: $ \av{\sigma_k^\mu a} \rightarrow J^\mu A $, $ \av{\sigma_k^\mu \sigma_k^\nu} \rightarrow  J^\mu J^\nu $ (with $ \mu\neq\nu $), and $ \av{a^\dagger a} \rightarrow |A|^2=X^2+P^2 $. Stationary solutions are then obtained by setting the left-hand side of Eq.~\eqref{eq:MFEoM} to $ 0 $. In order to make our analysis independent of the number of ions in the chain, we set $ \omega=\Omega N $.

For context, we briefly recall here the physics of the conventional (i.e., closed) version of the Dicke model, which has been introduced as a paradigmatic model to investigate the collective behavior of systems of spins coupled to bosonic degrees of freedom~\cite{Dicke:1954,Wang:1973,Hepp:1973,Kirton:2018}. It features a critical value of the spin-boson coupling associated with a quantum phase transition between a normal and a superradiant phase. In the latter, the bosonic harmonic oscillators show a finite displacement $ X $ from their equilibrium position. In the absence of inter-ion interactions (i.e., $ \mathcal{V}=0$) and state-dependent trapping potential (i.e., $ \omega_a=0 $), the generalization of the Dicke model to a dissipative environment, described by Eq.~\eqref{eq:QME}, is again characterized by the competition between two main phases. The bright phase is governed by the interplay between driving and decay and features a vanishing value of $ X $, while in the dark one the spin-phonon coupling suppresses both driving and dissipation and leads to a finite displacement of the CM mode, i.e. $ X>0 $. In contrast to the closed case, these phases coexist in a finite region of parameter space, resulting in intermittency in the fluorescence signal of the ions~\cite{Genway:2014}. 

In a trapped Rydberg ion simulator, new dynamical regimes emerge in the phase diagram as a consequence of the Rydberg ion-ion interaction and of the state-dependent trapping potential and, also, due to their interplay. Starting from the semi-classical EoMs of Eq.~\eqref{eq:MFEoM}, we will show that it exhibits a novel interaction-induced coexistence region, PL behavior, and multi-phase coexistence regimes. To benchmark our MF results we will numerically investigate the behavior of single quantum-jump Monte Carlo (QJMC) trajectories~\cite{Dalibard:1992,*Molmer:1993,*Plenio:1998}. These simulations provide access to the real-time dynamics of the photons emitted by the ions (see Fig.~\ref{fig:setup}), which can be detected in state-of-the-art experimental setups, and, therefore, allow to infer the various dynamical phases of the system. In the following, we will address systems with $ N $ ranging from $ 3 $ to $ 6 $, a cut-off on
the phonon Fock state up to $ N_\mathrm{ph} $ phonons and, for the sake of simplicity, with all-to-all coupling $ V_{kp}\approx V_0 \ \forall k,p$ between the ions~\cite{Note1}.

\textit{Interaction-induced coexistence region.---} To understand the effects of the interaction between ions, we begin by analyzing the simplest case with $ \Delta=0 $ and $ \omega_a=0 $. From Eq.~\eqref{eq:MFEoM} we obtain that, in the stationary state, $ J_z $ satisfies the following polynomial equation
\begin{equation}\label{eq:V:Jzequation}
\mathcal{A}^2 J_z^3 + \mathcal{A}^2 J_z^2 + (2\Omega+\mathcal{B}) J_z + \mathcal{B}=0,
\end{equation}
with 
\begin{equation}\label{eq:V:AandB}
\mathcal{A}=\frac{2}{\sqrt{\Omega}}\left(\mathcal V-\frac{2g^2 N}{\omega}\right)\quad \text{and} \quad \mathcal{B}=\frac{\gamma^2}{4\Omega}.
\end{equation}
Being a cubic equation in $ J_z $, Eq.~\eqref{eq:V:Jzequation} admits either one or three real solutions, depending on system parameters. In the latter case, whenever three real solutions coincide, i.e., when Eq.~\eqref{eq:V:Jzequation} takes the form $ (J_z-J_z^c)^3=0 $, a critical point emerges. In the $ g-\gamma $ plane, this occurs at
\begin{equation}\label{eq:V:criticalpoints}
\left(g^{\pm}_c,\gamma^{\pm}_c\right)= \left(\sqrt{\frac{\omega \Omega}{2N}}\sqrt{\frac{\mathcal{V}}{\Omega}\pm \frac{\sqrt{27}}{4}},\Omega\right),
\end{equation}  
with the critical point $ (g_c^-,\gamma_c^-) $ present if and only if the MF interaction $\mathcal{V}$ is larger than a threshold value $\mathcal{V}_\mathrm{thr}=\sqrt{27}\Omega/4 $. Therefore, the system has one (two) critical point (points) for $  \mathcal{V}<\mathcal{V}_\mathrm{thr} $  ($ \mathcal{V}>\mathcal{V}_\mathrm{thr} $). 

\begin{figure}
	\centering
	\includegraphics[width=\columnwidth]{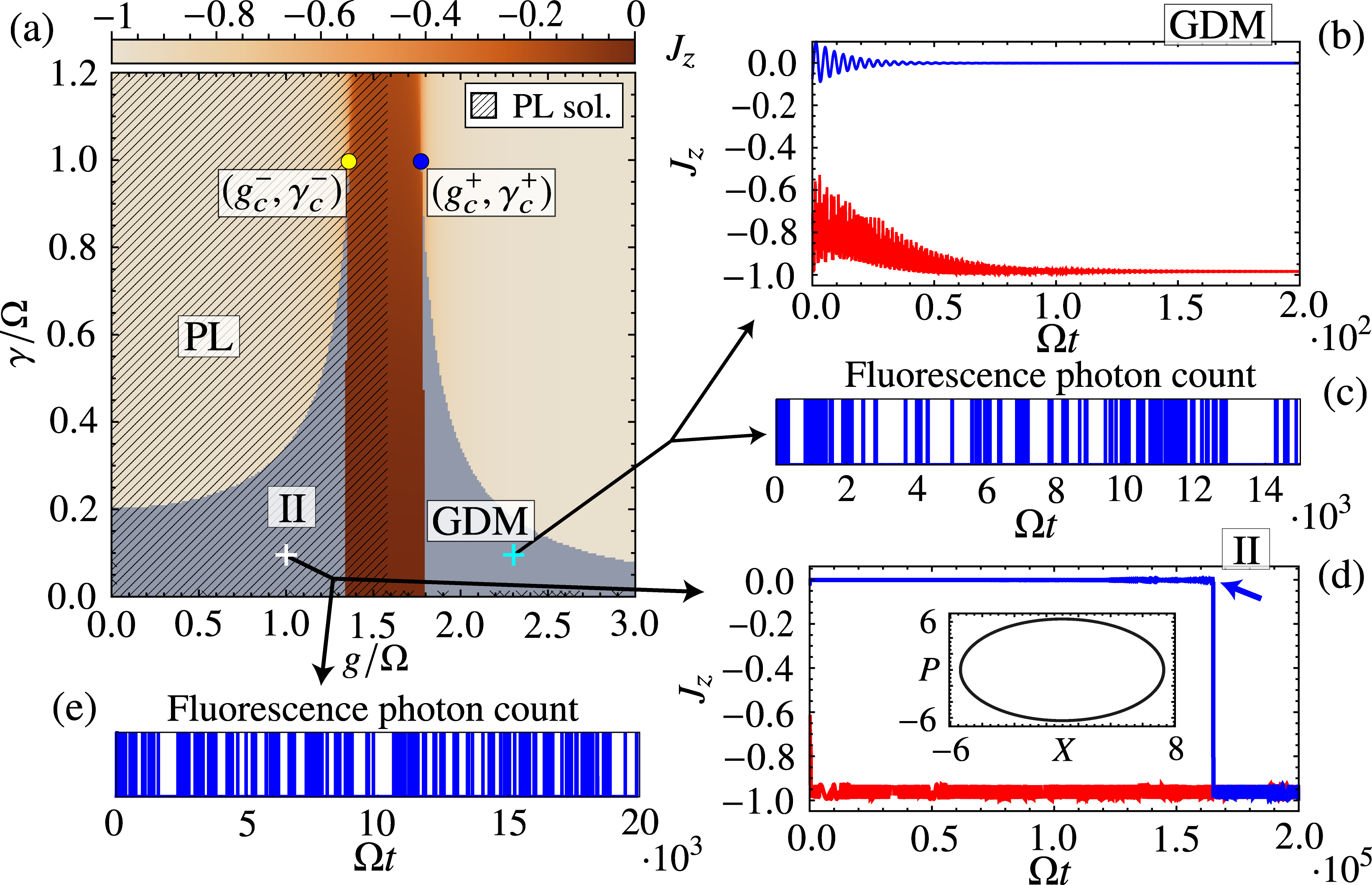}
	\caption{\textbf{Interaction-induced (II) coexistence region.} \textbf{(a)} Phase diagram in the $ g-\gamma $ plane for $ \mathcal{V}=5\Omega $, $ \Delta=0 $, and $ \omega_a=0 $. Gray areas denote the coexistence regions [3 real solutions to Eq.~\eqref{eq:MFEoM}], while the hatched area signals a PL regime. The II region originates as a consequence of inter-ion interactions and is present for $ \mathcal{V}>\mathcal{V}_\mathrm{thr} $, only. \textbf{(b, d)} Examples of the two possible MF dynamics of $J_z(t) $ as a function of time starting from neighborhoods of the different real solutions to Eq.~\eqref{eq:MFEoM} for $ \mathcal{V}=5\Omega $, $ \Delta=0 $, $ \omega_a=0 $, $ N=10 $, and \textbf{(b)} $ (g,\gamma)=(2.3,0.1) $ and \textbf{(d)} $ (g,\gamma)=(1,0.1) $. Here, a PL solution (red curve), with small amplitude oscillations of $ J_z(t) $ around a stationary value, emerges. Phase coexistence is present at small and intermediate times (see blue line and arrow). \textbf{Inset of (d):} Long-time limit-cycle dynamics of the CM mode position $ X(t) $ and momentum $ P(t) $, shown over a time window $ \Delta T=50\Omega^{-1} $ for the lasing solution. \textbf{(c, e)} Fluorescence photon count as a function of time in QJMC trajectories for $N=3$ ions with \textbf{(e)} and without \textbf{(c)} PL. Features of the MF solutions are clearly visible. Due to the emergence of PL, in the simulation we have considered $N_{\text{ph}}=400$ single-phonon states.}
	\label{fig:PDV5}
\end{figure}
Figure~\ref{fig:PDV5}(a) shows the phase diagram of the interacting model in the $ g-\gamma $ plane for $ \mathcal{V}=5 \Omega $. Given $ \mathcal{V}>\mathcal{V}_\mathrm{thr} $, two coexistence regions (gray areas), associated with the critical points $ (g_c^\pm,\gamma_c^\pm) $, emerge. From Eq.~\eqref{eq:V:criticalpoints}, we notice that, when both critical points are present, one always has $ g_c^- < g_c^+ $. Moreover, it is also possible to show that, within the MF treatment, the two coexistence regions never overlap. In particular, the region associated with $ (g_c^+,\gamma_c^+) $ is directly connected with the one found in the GDM discussed in Ref.~\cite{Genway:2014} and, therefore, we will refer to it as GDM region. Here, Eq.~\eqref{eq:V:Jzequation} has two stable solutions (corresponding to a bright and a dark phase, respectively) and an unstable one. This can be seen in Fig.~\ref{fig:PDV5}(b), where we show the ion magnetization $ J_z(t) $ for two different initial conditions. After a short transient, the system relaxes to two different stationary solutions, depending on the choice of the initial state. This behavior is also reflected in the fluorescence signal of photons emitted by the ions. Indeed, as can be seen from Fig.~\ref{fig:PDV5}(c), a typical time record of the fluorescence photon emission shows intermittency, i.e. alternating bright and dark periods. On the other hand, the region associated with $ (g_c^-,\gamma_c^-) $ emerges entirely as consequence of inter-ion interactions and it will thus be called interaction-induced coexistence region. As can be seen from Fig.~\ref{fig:PDV5}(a), where it is denoted by II, its nature is profoundly different from the one of the GDM region. Finite ion-ion interactions result in the emergence of a Hopf bifurcation: the stationary solutions to Eq.~\eqref{eq:MFEoM} become unstable and a limit-cycle behavior arises~\cite{Strogatz:2014}. This phenomenon, which manifests as self-sustained periodic oscillations in time, is the hallmark of a PL regime~\cite{Vahala:2009,Genway:2014}. This kind of behavior is seen in Fig.~\ref{fig:PDV5}(d): the stationary value of $J_z(t)$ displays fast small amplitude oscillations around $ J_z\approx-0.95 $ and the motion of $ X(t) $ and $ P(t) $ is clearly periodic (see inset). Interestingly, the presence of more than one real solution in the interaction-induced region [see the gray area II in Fig.~\ref{fig:PDV5}(a)] results in metastable behavior: two different initial conditions reach the same oscillating steady-state, but the timescale might be extremely long, so that effectively two phases coexist in the transient. For small systems, as considered here, this results in intermittent photon emission records [see Fig.~\ref{fig:PDV5}(e)].

\begin{figure}
	\centering
	\includegraphics[width=\columnwidth]{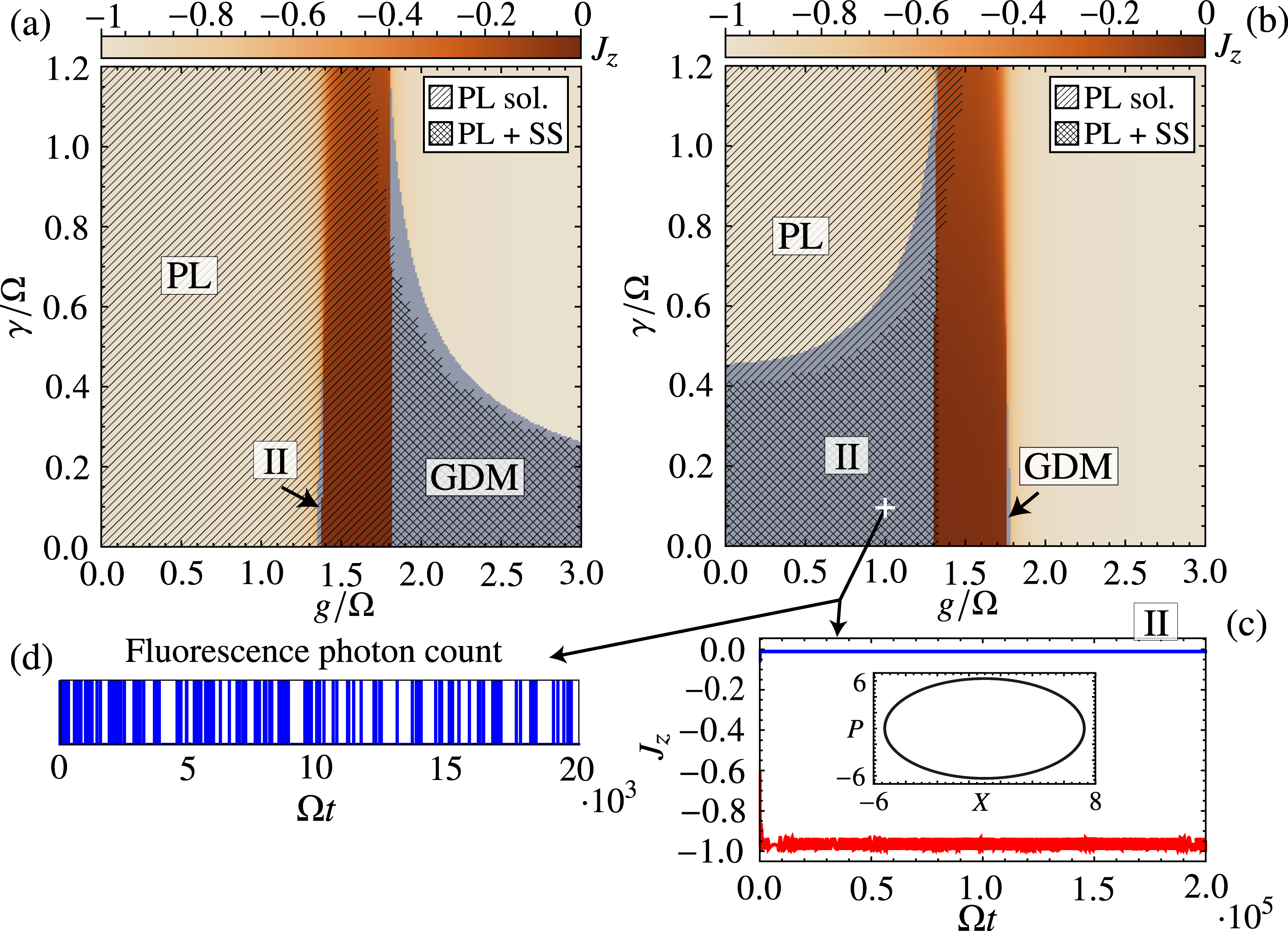}
	\caption{\textbf{Effects of a finite detuning.} Same as Fig.~\ref{fig:PDV5} for \textbf{(a)} $ \Delta=-0.1\Omega $ and \textbf{(b)} $ \Delta=0.1\Omega$. In the interaction-induced (II) region a PL and a stable solution (SS) coexist. \textbf{(c)} Same as Fig.~\ref{fig:PDV5}(b) for $ \Delta=0.1\Omega $. \textbf{(d)} Fluorescence photon count as a function of time in a QJMC trajectory for $  N=3 $ ions in the II coexistence region with $ N_\mathrm{ph}=400 $.}
	\label{fig:PDV5D}
\end{figure}

\textit{Finite detuning.---}
For a non-zero detuning $ \Delta $, Eq.~\eqref{eq:V:Jzequation} becomes
\begin{multline}\label{eq:V:JzequationD}
\mathcal{A}^2 J_z^3 + \left(\mathcal{A}^2+2\bar{\Delta}\mathcal{A}\right) J_z^2\\
 + (\bar{\Delta}^2+2\bar{\Delta}\mathcal{A}+2\Omega+\mathcal{B}) J_z + \left(\mathcal{B}+\bar{\Delta}^2\right)=0,
\end{multline}
with $ \mathcal{A} $ and $ \mathcal{B} $ given in Eq.~\eqref{eq:V:AandB}, and $ \bar{\Delta}=2\Delta/\sqrt{\Omega} $. Equation~\eqref{eq:V:JzequationD} is, again, a cubic equation in $ J_z $. Critical points $ (g_c,\gamma_c) $ arise when three coinciding real solutions are present. This occurs when the following equations are satisfied 
\begin{equation}\label{key}
\begin{cases}
\left(\mathcal{A}-\bar{\Delta}\right)^3-\frac{27}{4}\Omega\mathcal{A}=0,\\
\mathcal{B}-\frac{1}{3}(\mathcal{A}+\bar{\Delta})^2+2\Omega=0.
\end{cases}
\end{equation} 
The number of critical points can be obtained by inspecting the determinant of Eq.~\eqref{eq:V:JzequationD}. As discussed in Ref.~\cite{Genway:2014}, in contrast to the closed version of the Dicke model~\cite{Emary:2004}, a small but finite detuning $ \Delta $ does not alter significantly the phase diagram shown in Fig.~\ref{fig:PDV5D}, i.e., it does not destroy the dynamical phase transition discussed previously. However, for larger values of $ \Delta $ the behavior of the phase diagram strongly depends on its sign. In particular, when $ \Delta<0 $ the interaction-induced region is drastically suppressed while the GDM one is enhanced. Moreover, the region at small $ g $ subject to Hopf instability extends to a larger portion of the phase diagram [see the hatched area in panel (a)]~\cite{Genway:2014}. The behavior of the system for $ \Delta>0 $ is even more interesting. Here, besides an overall suppression of the GDM coexistence region and an enhancement of the interaction-induced one, the detuning significantly affects the stability of steady state solutions. As can be seen in Fig.~\ref{fig:PDV5D}(b, c), for small $ g $ and $ \gamma $ a region with a stable, non-oscillating solution emerges within the PL regime [compare with Fig.~\ref{fig:PDV5}(a,d)]. 

So far we have analyzed, within a MF approach, the different non-equilibrium regimes induced by ion-ion interactions. The latter is responsible for the emergence of a lasing regime and a novel coexistence region for small spin-phonon coupling $ g $. A finite detuning $ \Delta $, suppressing the interaction-induced coexistence region ($ \Delta<0 $) or the GDM one ($ \Delta>0 $), allows to modify the phase diagram of the system and to control its stability property. In all these cases, phase coexistence results in intermittency in the fluorescence signal of photons emitted by the ions [see Figs.~\ref{fig:PDV5}(c,e) and Fig.~\ref{fig:PDV5D}(d)].

\begin{figure}
	\centering
	\includegraphics[width=\columnwidth]{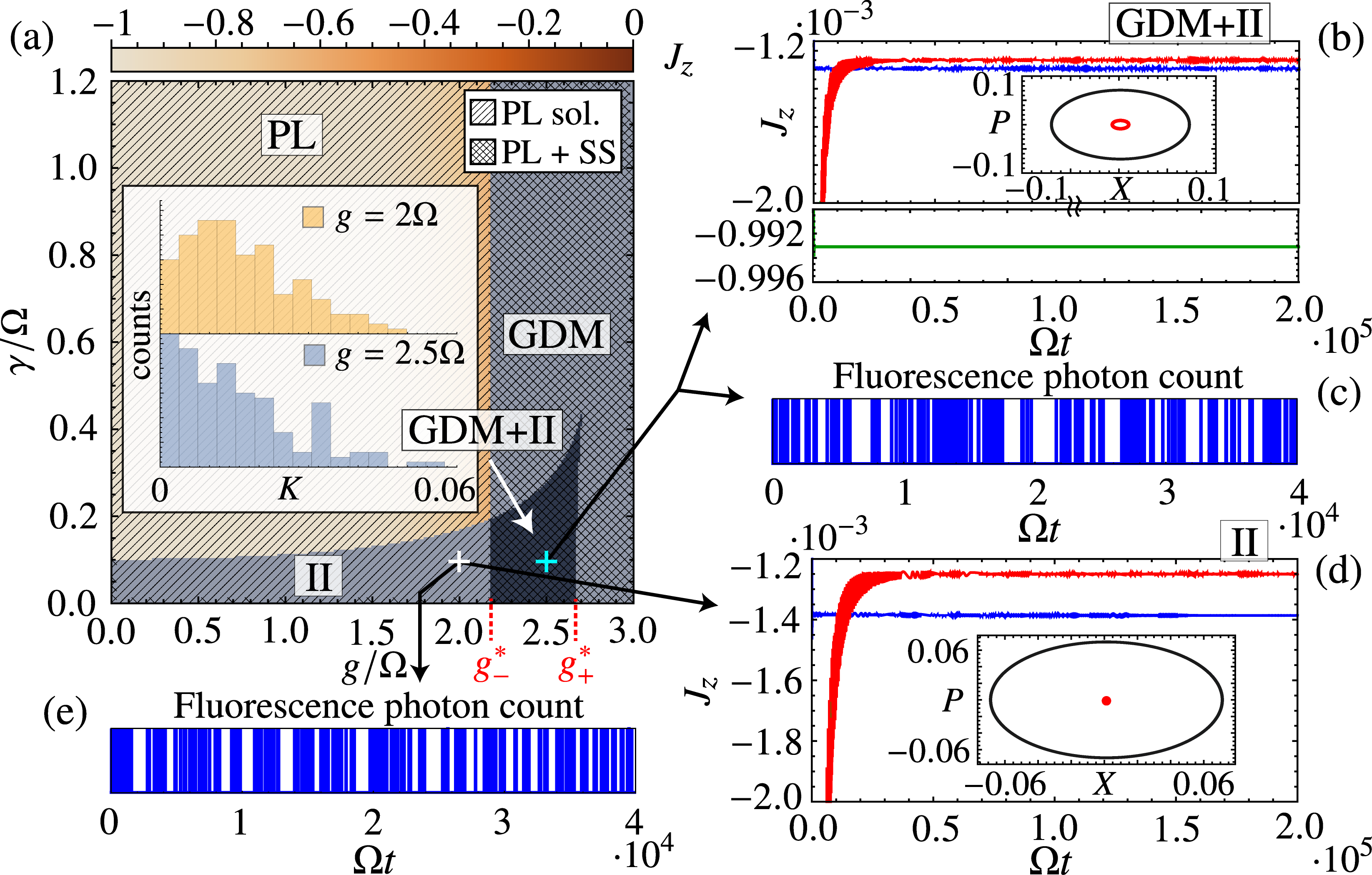}
	\caption{\textbf{Emergence of a multi-coexistence regime for $ \omega_a\neq 0 $. } \textbf{(a)} Same as Fig.~\ref{fig:PDV5} for $ \mathcal{V}=10\Omega $, and $ \omega_a=2\Omega $. The dark gray area GDM$ + $II denotes a multi-coexistence region with 5 real solutions to Eq.~\eqref{eq:MFEoM}, while in the interaction-induced (II) region PL and stable solutions coexist. \textbf{Inset:} Distributions of the rate $ K $ of emitted photons over a time window of $ \Delta t=10^3\Omega $ for the photon count records of panels (e) (top, yellow) and (c) (bottom, blue). The two histograms show qualitative differences hinting, indeed, towards a different character of the stationary state (see text). \textbf{(b, d)} Same as Fig.~\ref{fig:PDV5}(b) for $ \mathcal{V}=10\Omega $, $ \omega_a=2\Omega $, and  \textbf{(b)} $ (g,\gamma)=(2.5,0.1) $ and \textbf{(d)} $ (g,\gamma)=(2,0.1) $ . In both panels two lasing solutions (red and blue lines) coexist, while in the GDM$ + $ II phase an additional stable solution (green line) is present [panel (b)]. \textbf{Insets:} Long-time limit-cycle dynamics of the phononic position $ X(t) $ and momentum $ P(t) $ over a time window $ \Delta T=50\Omega^{-1} $ for the two lasing solutions of the main panels. \textbf{(c, e)} Fluorescence photon count as a function of time in QJMC trajectories inside the \textbf{(e)} II and \textbf{(c)} GDM$ + $II regions for $ N_\mathrm{ions}=6 $ and $ N_\mathrm{ph}=100 $.}
	\label{fig:PDV10D}
\end{figure}

\textit{State-dependent trapping potential.---}
We now turn to the case with both ion-ion interactions and a (strong) state-dependent trapping potential $ \omega_a $, which is of great relevance for Rydberg trapped ion quantum simulators. Here, fixed points of Eq.~\eqref{eq:MFEoM} satisfy a $ 7$th order polynomial equation,
\begin{equation}\label{eq:oma:Jzequation}
\sum_{j=0}^7 c_j J_z^j=0,
\end{equation}
where coefficients $ c_j $, which depend on system parameters, are too cumbersome to be reported here. Being the analytical investigation of critical points particularly involved, we resort here to a numerical analysis only. The latter reveals that, at least for not too strong spin-phonon coupling $ g $ and atomic decay $ \gamma $, only two $ 3 $rd order critical points emerge. Interestingly, as can be seen in Fig.~\ref{fig:PDV10D}(a), the presence of a state-dependent trapping potential $ \omega_a $ is responsible for the occurrence of a multi-coexistence regime between $ g^{*}_- $ and $ g^{*}_+ $, in which the GDM and the interaction-induced coexistence regions merge [dark gray area GDM$ + $II  in panel (a)]. Here, Eq.~\eqref{eq:V:JzequationD} possesses five real solutions. As shown in Figs.~\ref{fig:PDV10D}(a,b), one solution is stable while the remaining ones give rise to two different limit-cycle solutions.  On the other hand, it also emerges that in the interaction-induced region, i.e.,  for $ g<g^{*}_- $, no stable solution is present and two lasing solutions coexist. 

Fingerprints of these different regimes can be found by analyzing the distributions of the rate $ K $ of photon emitted by the ions over a fixed time window ~\cite{Garrahan:2010,Ates:2012}. Although both emission records shown in Figs.~\ref{fig:PDV10D}(c,e) display intermittent behavior, the distribution of photons emitted is different in both cases. The corresponding histograms are shown in the inset of Fig.~\ref{fig:PDV10D}(a): the distribution of $ K $ inside the multi-coexistence regime shows an enhanced value near $ K = 0 $ and a higher peak at larger values of $ K $ with respect to the interaction-induced region. This indicates, indeed, a qualitative change of the stationary state as anticipated from the MF analysis, i.e., it hints towards the existence of a multi-stable regime.

\textit{Conclusions.---} 
In this work we have investigated a plethora of dynamical regimes realized in a trapped Rydberg ion quantum simulator. We have shown how the interplay between electron-phonon coupling and dipolar ion-ion interactions results in a rich dynamical behavior. Here, a newly interaction-induced phase coexistence region emerges and multi-phase coexistence can be achieved. Moreover, the presence of finite dipolar interactions gives rise to a region with stable limit-cycle solutions, associated with phonon lasing. Signatures of the different dynamical regimes can be detected through the time-resolved spectroscopy of emitted photons. 

\begin{acknowledgments}	
The research leading to these results has received funding from the  European  Research Council under the European Union's Seventh Framework Programme (FP/2007-2013)/ERC Grant Agreement No.~335266 (“ESCQUMA”), the EPSRC Grant No. EP/M014266/1, and the EPSRC Grant No. EP/R04340X/1 via the QuantERA project “ERyQSenS”. F. M. G. acknowledges the hospitality of the Stockholm University and would like to thank M. Hennrich, G. Higgins, F. Pokorny, and C. Zhang for inspiring discussions.  I.L.~gratefully acknowledges funding through the Royal Society Wolfson Research Merit Award.
\end{acknowledgments}

\footnotetext[1]{Within the system sizes we have considered, simulations with a local interaction potential of the form $ V_{kp}=C_3/|\bm{r}_k-\bm{r}_p|^{3} $ give the same qualitative results although require more numerical resources.}

\bibliography{bibliography.bib}

\end{document}